\documentstyle[twoside,fleqn,espcrc2]{article}

\newcommand{\AmS}{{\protect\the\textfont2
  A\kern-.1667em\lower.5ex\hbox{M}\kern-.125emS}}

\title{Strings with Extrinsic Curvature: An Analysis of
the Crossover Regime\thanks{presented by G. Harris}
}
\author{Mark Bowick, Paul Coddington, Leping Han, Geoffrey Harris
\address{Dept. of Physics and NPAC, Syracuse
        University\\
        Syracuse, NY 13244, USA}
and Enzo Marinari\address{Dept. of Physics and NPAC, Syracuse
University and
Universit\'a di Roma {\it Tor Vergata},
Viale della Ricerca Scientifica, 00173 Roma, Italy}}

\begin{document}

\begin{abstract}
We present the results of a set of Monte Carlo simulations of Dynamically
Triangulated Random Surfaces embedded in three dimensions with an extrinsic
curvature dependent action.  We analyze several
observables in the crossover regime and discuss whether or not our
observations are indicative of the presence of a phase transition.
\end{abstract}

\maketitle
\begin{flushright}
{\bf SU-HEP-4241-527},
{\bf hep-lat/9211024}.
\end{flushright}
\section{Introduction}

In this work, we investigate a theory of fluid,
fluctuating random surfaces embedded
in three dimensions.  Theories of fluctuating surfaces (string theories)
have been conjectured to describe a wide variety of physical phenomena and
models, including the strong interaction at large distances, the
3d Ising model, and unified theories incorporating gravity.
Lipid bilayers and microemulsions are also examples of fluctuating surfaces
\cite{DAVID}.  Such
biological and chemical membranes exhibit self-avoidance,
which we do not take into account in our simulations.  Models of
fluctuating random surfaces can in fact be solved exactly when
the surfaces are embedded in dimensions $D \leq 1$; these solutions
break down though when continued to the more physical regime $D > 1$.
This may be related to the observation that simulations
of fluctuating surfaces in $D >1$ using a discretization
of the standard Polyakov string action are dominated by crumpled,
spiky configurations.

    Our lattice model is constructed by triangulating each surface.
Each node of the triangulation is embedded in $R^{3}$ by the functions
$X_i^{\mu}$; $i$ labels the $i$th node and $\mu$ runs from $1$ to $3$.
The triangulation is characterized by the adjacency matrix $C_{ij}$,
whose elements equal $1$ if $i$ and $j$ label neighboring nodes of
the triangulation and vanish otherwise.
The triangles are
assumed to be equilateral (as measured by the intrinsic metric
of each surface); the connectivity at each node determines the
intrinsic curvature.
We also associate a normal
vector $(n^{\mu})_{\hat{k}}$ with each triangle (indices with hats
label the triangles).  We shall study the theory defined by the action
\cite{CAT,BJW,BCJW,CKR,A1,A2,US}
\begin{eqnarray}
\protect\label{ourdisact}
        & \nonumber S =  S_{G} + \lambda S_E =
    \sum_{i,j,\mu}C_{ij}(X^{\mu}_i - X^{\mu}_j)^2 + \\ &
 \lambda\sum_{\hat{k},\hat
{l},\mu}C^{\hat{k}\hat{l}}(1 -
  n^{\mu}_{\hat{k}}\cdot n^{\mu}_{\hat{l}})\ .
\end{eqnarray}
For $\lambda = 0$ this is simply a discretization of the Polyakov string
action.  The final term, which depends on the discretized extrinsic
curvature, introduces a ferromagnetic interaction
between surface normals, which one might hope would cause smoother
surfaces to dominate the partition function.
We would like
to know if there is a smooth phase and a phase transition (at
some finite $\lambda_c$) between this phase and the
crumpled phase observed
at $\lambda = 0$.  If this were so, an interesting
continuum limit of this lattice model could perhaps be constructed
at this phase transition point, yielding a new continuum string theory.


\section{The Simulation}
     We have considered triangulations with the topology of the torus,
to minimize finite size effects.
The above action  was used, with
the BRST invariant measure utilized also by Baillie, Johnston and Williams
\cite{BJW}, so that

\begin{equation}
  Z = \sum_{G \in T(1)}
  \int\prod_{\mu,i}dX^{\mu}_{i}\prod_{i}q_i^{\frac{D}{2}}
  \exp( -S_{G} - \lambda S_E)\ ,
\end{equation}
where $D=3$, $q_i$ is the connectivity of the $i$th vertex and $T(1)$ refers
to the set of triangulations of genus $1$.  We used the standard Metropolis
algorithm to update our configurations.
To sweep through the space of
triangulations we performed flips on randomly
chosen links.  Flips were automatically rejected if they yielded a degenerate
triangulation.
After a set of  $3M$ flips was
performed, $3M$ randomly selected embedding coordinates were updated via
random shifts from a flat distribution.
Most of the  Monte Carlo simulations were performed on
HP-9000 (720 and 750 series) workstations; we also collected some data by
simulating lattices on each of the 32 nodes of a CM-5.


%

We ran on lattices ranging in size from $N = 36$ to $576$ ($N$ signifies
the number of vertices) with $4$ to $7$ different values
of $\lambda$ for each $N$.  Most of the data was this data was taken in the
region $\lambda \in (1.325,1.475)$.  For small $N$, the runs consisted
of $3 \times 10^6$ sweeps, while we performed longer runs (of up to
$27 \times 10^6$ sweeps for $N=576$) for larger lattices,
because the auto-correlation
times for our simulations were very large.  (The correlation time
for the radius of gyration was greater than $10^6$ sweeps
for $N=576$!)  To determine our
observables as a function of $\lambda$ we used a histogram
reconstruction procedure.  We patched different histograms by
weighting them with the associated statistical indetermination
(which was estimated by a jack-knife binned procedure).  Various consistency
checks indicate that this procedure is very reliable.
\section{Observables}
     We measured the edge action $S_E$ and the associated specific heat
$C(\lambda) \equiv \frac{\lambda^2}{N}(<S_E^2> - <S_E>^2)$.   In Fig. 1
we plot the specific heat curve (constructed via the histogram procedure)
and we tabulate its maximum and peak position for various lattice sizes
in Table $1$.
\begin{figure}[htb]
\makebox[55mm]{\rule[-21mm]{0mm}{43mm}}
\caption{$C(\lambda)$. Dotted lines: N=144. Dashed lines: N=288.
Solid lines: N=576.}
\label{fig:largenenough}
\end{figure}
\begin{table}
\begin{tabular}{|l|l|l|} \hline
N & $C(\lambda)^{\rm{max}}$ & $\lambda_c$ \\ \hline
36 & 3.484(8) & 1.425(35) \\ \hline
72 & 4.571(15) & 1.410(15) \\ \hline
144 & 5.37(14) & 1.395(30) \\ \hline
288 & 5.55(7) & 1.410(25) \\ \hline
576 & 5.81(17) & 1.425(30) \\ \hline
\end{tabular}
\protect\caption[CT_TWO]{The maximum of the specific heat and its position,
with errors, for different lattice sizes. \protect\label{T_TWO}}
\end{table}

     We see that the specific heat peak grows vigorously with $N$ for small
lattices, but that this growth quickly levels off for larger $N$. These
observations agree fairly well with previous work \cite{CKR,A1,A2}. For the
larger lattices it appears that the peak position shifts very slowly
towards higher values of $\lambda$, though this increase is not
statistically significant.  The shape of the peak does not change dramatically
with $N$; it narrows perhaps a bit between $N=144$ and $N=576$.

     We measured the magnitude of the extrinsic Gaussian curvature,
$\int \mid K \mid \sqrt{\mid h \mid}$ ($h$ is the induced metric and
$K$ is the determinant of the extrinsic curvature matrix), given by
  \begin{equation}
    \mid {\cal{K}} \mid = \frac{1}{N}\sum_i\mid 2\pi -
    \sum_{\hat{j}} \phi_i^{\hat{j}}\mid\ .
    \protect\label{E_CK}
  \end{equation}
Here $\phi_i^{\hat{j}}$ denotes the angle subtended by the $\hat{j}$th
triangle at the $i$th vertex.  This quantity, plotted in Fig. 2,
measures the
magnitude of the deficit angle in the embedding space averaged over
all vertices.
\begin{figure}[htb]
\makebox[55mm]{\rule[-21mm]{0mm}{43mm}}
\caption{$\mid {\cal{K}} \mid$, plotted as in Figure $1$.}
\label{fig:smallenough}
\end{figure}
Note that the mean Gaussian curvature decreases rapidly
in the neighborhood of $\lambda = 1.4$, indicating that a sharp
crossover is occuring in this system.  From this plot we can see that
finite size effects increase with $\lambda$; they do not appear
to peak in the region about $\lambda = 1.4$ as one might expect
for a typical phase transition.
We also measured various other observables which
characterize both the intrinsic and extrinsic geometry of these surfaces
(and the correlation between intrinsic and extrinsic geometry).  These
measurements are discussed in a longer write-up of this work
\cite {US}.  They all exhibit sharp crossover behavior in
the region near $\lambda= 1.4$.  We found that the auto-correlation times
of these observables grew rapidly as $\lambda$ increased, but we did not
note any maximum in these times in the region about $\lambda = 1.4$.
\section{Interpretation}
This model of crumpled surfaces appears to
exhibit sharp crossover behavior in the region around $\lambda = 1.4$.  The
sharp change in the magnitude of the Gaussian curvature,
the radius of gyration and other observables
indicates that the normals acquire long-range correlations, up to
the size of the systems we examine.  The zero string tension
measurement of
\cite{A2} also shows that the disordered regime differs from the regime in
which the surfaces are ordered (up to scale of the lattices that are
simulated) by only  a small shift in $\lambda$. This evidence  might indicate
the presence of a phase transition at this point.  Since the peak
growth rapidly diminishes for large $N$, such a phase transition would
likely be higher than second order.
Still, the apparent absence of
diverging correlation times and increasing finite size effects in the
peak region leads us to question whether we are actually observing
a typical phase transition.

There are indeed other possible interpretations of our data.
Note that the surfaces which we
simulate are quite small.
For instance, if the surfaces in our simulations had an
intrinsic dimension of $2.87$ (characteristic of $D=0$ gravity),
they would have roughly a linear size of fewer than $10$
lattice spacings.

Perhaps the simplest alternative explanation  for the presence of this peak is
suggested by the arguments of Kroll and Gompper \cite{KROGOM}.  They argue
that the peak occurs when the persistence length of the system approaches the
size of the lattice ($\xi_p \sim N^{\frac{1}{d}}$); $d$ denotes the intrinsic
Hausdorff dimension.
Fluctuations on a larger scale become more
important.  When this scale is greater than the lattice size these
fluctuations are suppressed. Thus one might surmise that the specific heat
will drop for large $\lambda$.
                          The one-loop renormalization group calculation
\cite {RG} predicts that
the persistence length grows as $\xi_p \sim \exp(C\lambda)$; $C$ is inversely
proportional to the leading coefficient of the  beta function.  We would
expect that the peak position should shift  to the right with increasing $N$
in this scenario as
\begin{equation}
  \lambda_{peak}(N') - \lambda_{peak}(N) =
\frac{\ln(\frac{N'}{N})}{dC}.
\end{equation}
This reasoning also indicates that the peak should widen as the lattice size
increases; we do not observe this at all.

      An alternative scenario, which builds on the ideas in the above
paragraph, is suggested by the tantalizing similarities between the results of
our fluid surface simulations and what has been observed for the $d=4$ $SU(2)$
Lattice Gauge Theory \cite{ENZOHI} and for the $d=2$ $O(3)$ model.

The $O(3)$ model, which is thought to be
asymptotically free, exhibits a specific heat peak
near $\beta = 1.4$ (first measured via Monte Carlo
simulations by Colot \cite{COLOT}).  The
origin of this peak is understood \cite{BROUT}; it is due to the
fluctuations of the sigma particle, a low-mass bound state of the massless
$O(3)$ pions.  The sigma induces short-range order and contributes to the
specific heat as a degree of freedom only at high temperatures (when the
correlation length in the system becomes smaller than its inverse mass).
The peak thus occurs at the beginning of the crossover regime, when the
correlation length is several lattice spacings.

According to the low temperature expansion, the correlation  length grows as
$\xi \sim \exp(2\pi\beta)/\beta$. Thus one would expect  a fairly rapid
crossover in the $O(3)$ model;  the correlation length should increase by
roughly a factor of $9$ when $\beta$ is shifted by about $0.35$.
Such a crossover is indeed observed, though it
is not so apparent that it is as dramatic as the crossover behaviour
observed for fluid surfaces.

Recent simulations of the $O(3)$ model \cite{KOSTAS} indicate that the
specific heat peak grows significantly when the lattice size $L$ is increased
from $5$ to $15$, and that virtually no growth in peak height is evident as
$L$ is increased further up to $100$.  Furthermore, the
peak position shifts to the
right as $L$ grows and then appears to stabilize for large $L$.  This is more
or less what we observe in our simulations of fluid surfaces, on lattices of
small size.  We point out these similarities largely to emphasize that there
does exist an asymptotically free theory (with low mass excitations) which
exhibits crossover behavior qualitatively similar to that
observed in our simulations.
The analogy is perhaps deeper, though, since the fluid surface action (with
extrinsic curvature) in certain guises looks like a sigma model action.  It
would not therefore be so surprising from this point of view to find a sigma
particle in these theories, perhaps associated with ($\hat{n}^2 -1$), in which
$\hat{n}$ denotes the unit normal to our surfaces.

\section{Acknowledgments}
This work has been done with NPAC (Northeast Parallel Architcture Center)
and CRPC (Center for Research in Parallel Computing) computing facilities.
The research of MB was supported by the Department of Energy Outstanding
Junior Investigator Grant DOE DE-FG02-85ER40231 and that of GH by
research funds from Syracuse University.  We gratefully acknowledge
discussions, help and sympathy from Jan Ambj{\o}rn, Kostas
Anagnostopoulos, John Apostolakis, Clive Baillie, Mike Douglas, David
Edelsohn, Geoffrey Fox, Volyosha Kazakov, Emil Martinec, Alexander
Migdal, David Nelson, Giorgio Parisi, Bengt Petersson, Steve Shenker, and
Roy Williams.  Deborah Jones, Peter Crockett, Mark Levinson and Nancy
McCracken provided invaluable computational support.


\begin{thebibliography}{9}
  \bibitem{DAVID}
    See F. David, in {\em Two Dimensional Quantum Gravity
    and Random Surfaces}, eds. D.J. Gross, T. Piran, and S. Weinberg,
    World Scientific (1992) pp. 80-124.
  \bibitem{CAT}
    S. Catterall,
    Phys. Lett. {\bf 220B} (1989) 207.
  \bibitem{BJW}
    C. Baillie, D. Johnston and R. Williams,
    Nucl. Phys.  {\bf B335} (1990) 469.
  \bibitem{BCJW}
    C. Baillie, S. Catterall, D. Johnston and R. Williams,
    Nucl. Phys. {\bf B348} (1991) 543.
  \bibitem{CKR}
    S. Catterall, J. Kogut and R. Renken,
    Nucl. Phys. {\bf B} (Proc. Suppl.) {\bf 99A} (1991) 1.
  \bibitem{A1}
    J. Ambj{\o}rn, J. Jurkiewicz, S. Varsted, A. Irb\"{a}ck
    and B. Petersson,
    Phys. Lett. {\bf 275B} (1992) 295.
  \bibitem{A2}
    J. Ambj{\o}rn, A. Irb\"{a}ck, J. Jurkiewicz and B. Petersson,
        Niels Bohr Institute Preprint
     NBI-HE-92-40 (hep-lat/9207008).
  \bibitem{US}
    M. Bowick, P. Coddington, L. Han, G. Harris, and E. Marinari,
       Syracuse and Rome preprint
     SU-HEP-4241-517, SCCS-3557, ROM2F-92-48, hep-lat/9209020,
     submitted to Nucl. Phys. {\bf B}.
  \bibitem{KROGOM}
    D. Kroll and G. Gompper,
    Europhys. Lett. {\bf 19} (1992) 581;
  \bibitem{RG}
     A.M. Polyakov,
    Nucl. Phys. {\bf B268} (1986) 406 and
     H. Kleinert,
     Phys. Lett. {\bf 174B} (1986) 335.
  \bibitem{ENZOHI}
    M. Falcioni, E. Marinari, M. L. Paciello, G. Parisi and B. Taglienti,
    Phys. Lett. {\bf 102B} (1981) 270.
  \bibitem{COLOT}
    J. L. Colot,
    J. Phys. {\bf A16} (1983) 4423.
  \bibitem{BROUT}
    J. Orloff and R. Brout,
    Nucl. Phys. {\bf B270} (1986) 273.
  \bibitem{KOSTAS}
    K. Anagnostopoulos, private communication.
\end{thebibliography}
\end{document}